\documentstyle[epsf,bibnorm]{lamuphys}

\makeatletter
\let\chapter\hid@chapter
\makeatother
\begin{document}
\pagenumbering{arabic}

\def\lapprox{\mathrel{\mathop
  {\hbox{\lower0.5ex\hbox{$\sim$}\kern-0.8em\lower-0.7ex\hbox{$<$}}}}}
\def\gapprox{\mathrel{\mathop
  {\hbox{\lower0.5ex\hbox{$\sim$}\kern-0.8em\lower-0.7ex\hbox{$>$}}}}}

\title{Nuclear astrophysics and neutrinos}

\author{G. Fiorentini\inst{1,2} }

\institute{ Dipartimento di Fisica, Universit\`a di Ferrara,
Via Paradiso 12, I-44100 Ferrara, Italy
\and
Istituto Nazionale di Fisica Nucleare, Sezione di Ferrara,
Via Paradiso 12, I-44100 Ferrara, Italy  }

\maketitle

\noindent
{\small Talk presented
at ``VII Convegno su Problemi di Fisica Nucleare Teorica'',     
Cortona (Italy),  19-21 October 1998 }

\begin{abstract}
In this report, centered on the activities within the MURST-PRIN project  
``Fisica teorica del nucleo e dei sistemi a pi\`u corpi" we discuss recent 
advances on the following items:
i)neutrinos as  probes of the solar interior and of other astrophysical 
objects;
ii)neutrinos as  probes of physics beyond the Standard Model  of 
electroweak interactions;
iii)the role of nuclear physics in i) and ii);
iv) who is doing what within the italian network;
v)future projects.
\end{abstract}

\section{Introduction}

After the results of solar neutrino experiments in the last few years
(see for reviews \cite{gf:report,gf:BKS98}) and 
the  recent impressive data reported by Superkamiokande
\cite{gf:sk,gf:skatmos} on both solar and 
atmospheric neutrinos,
we are living a really exciting phase of neutrino physics.
Nuclear physics is deeply involved in it, since neutrinos are
produced  as a result of nuclear reactions and are detected in the 
laboratory generally  by means of nuclear interactions.

In this report, centered on the activities within the MURST-PRIN project  
``Fisica teorica del nucleo e dei sistemi a pi\`u corpi" we discuss recent 
advances on the following items:\\

\noindent
i)neutrinos as  probes of the solar interior, and of other astrophysical 
objects;\\

\noindent
ii)neutrinos as  probes of physics beyond the Standard Model  of 
electroweak interactions;\\

\noindent
iii)the role of nuclear physics in i) and ii);\\

\noindent
iv) who is doing what within the italian network;\\

\noindent
v)future projects.\\

With respect to the last item, we concentrate on the calculation of $hep$ 
and other rare neutrinos, we outline the interesting problems posed by a 
recent experimental proposal, LENS \cite{gf:lens},
 and the physics potential of a new  
undeground apparatus for nuclear astrophysics, LUNA2 \cite{gf:luna2},
 which will be 
installed at Gran Sasso in the next few years.

\section{Neutrino as  probes of astrophysical objects}
\label{secneutrino}

Due to their extremely  long mean free path, neutrinos are  ideal 
probes of stellar interiors. As an example, the heat and light we get from the 
sun correspond to photons generated in the outermost layer of the solar 
atmosphere  whereas neutrinos directly emerge 
from the solar core and can probe the innermost part  of the star, a 
region otherwise unaccessible to observations. In this section we briefly 
discuss just a few topics, so as to outline the potential of neutrinos as 
probes of  stellar interiors. Actually, the structure of other 
astrophysical objects can  also be determined by exploiting the long mean 
free path of neutrinos. As an example, we shall present a neutrino map of 
the Galaxy at the end of this section.

\subsection{Nuclear energy production in stars}
\label{subsecnuclear}

Since the pioneer papers in the thirties by Bethe \cite{gf:B39},
 Bethe and Critchfield \cite{gf:BC38}, 
concerning the role of  the pp chain and the CNO 
cycle,  theoretical calculations of nuclear energy production in stars 
have greatly advanced and have reached a high degree of complexity and 
sofistication. On the other hand, for the succeeding fifty years there was 
no real observation of the fact that  stellar energy  is 
generated by means of nuclear fusion.

In fact, no matter which is the detailed mechanism, Hydrogen burning 
requires neutrino emission due to (global) lepton number conservation:
\begin{equation}
\label{eq1}
4p+2e \rightarrow ^4He +2 \nu + {\mbox{heat}}\, ,
\end{equation}
and neutrinos are the only product of nuclear reactions which can 
escape, undisturbed, from the solar energy generating core. 

In the early nineties Gallex \cite{gf:gallex}
 and Sage \cite{gf:sage}, two experiments sensitive to  pp 
neutrinos (i.e. those from $p+p \rightarrow d + e +\nu_e$)
 have shown that 
the neutrino signal agrees, within a factor two,  with the  assumption 
that Eq. (\ref{eq1}) is the source of solar power.  

Conceptually this result, i.e. the observational proof that 
the sun is powered by 
nuclear fusion, is at least as important as the missing factor two, the so 
called solar neutrino puzzle.

\begin{figure}
\centerline{
\epsfysize=12cm \epsfbox{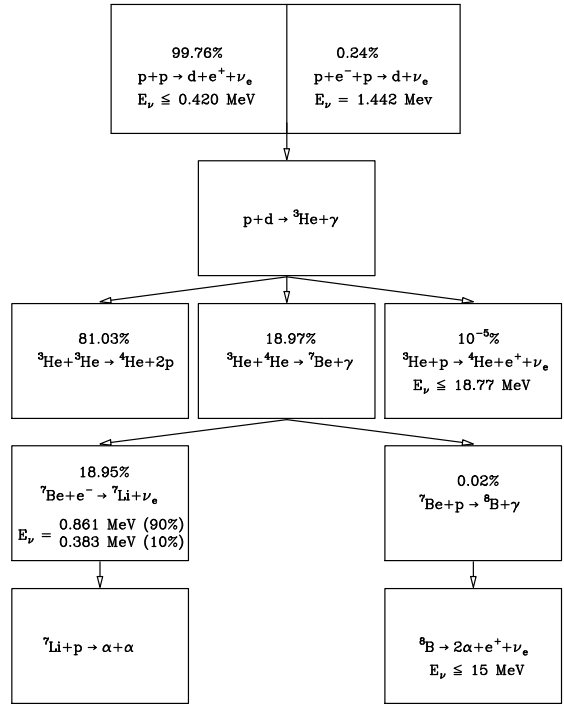}  
}
\caption[ccc]{The pp chain}
\label{figpp}
\end{figure}

\begin{figure}
\centerline{
\epsfysize=15cm \epsfxsize=10cm \epsfbox{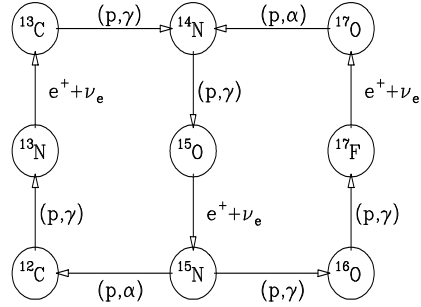}  
}
\vspace{-5cm}
\caption[ccc]{The CNO cycle}
\label{figcno}
\end{figure}

\begin{figure}
\centerline{
\epsfysize=10cm \epsfxsize=10cm \epsfbox{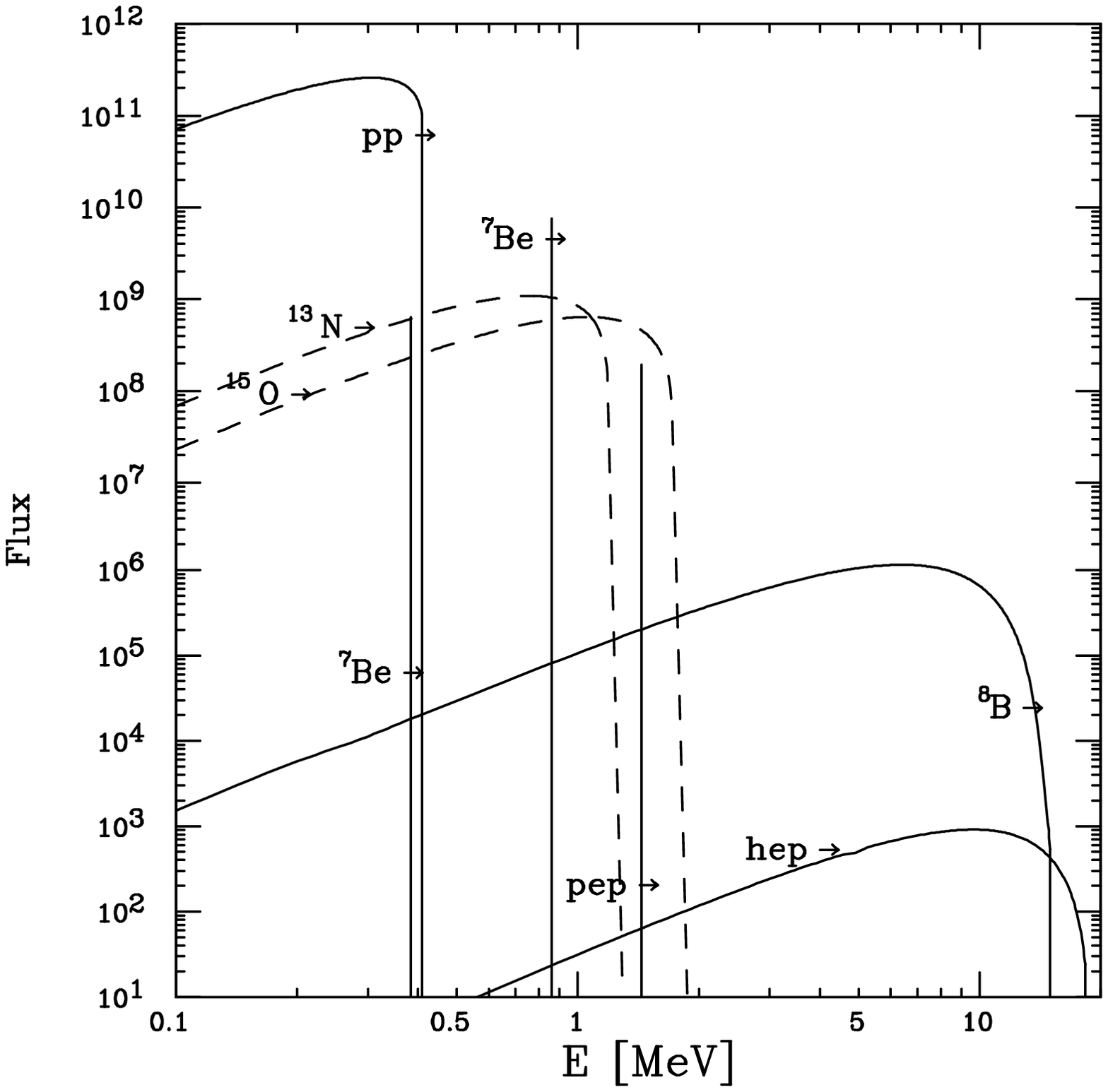}  
}
\caption[ccc]{The solar neutrino spectrum}
\label{figspe}
\end{figure}

Neutrinos from the different branches of Eq. (\ref{eq1}), 
see Figs. \ref{figpp}, \ref{figcno} and  \ref{figspe}, can be 
discriminated due to the different energies. As an example, pp neutrinos 
have a continuous spectrum up to $E_\nu$=0.420 MeV, whereas  $^7$Be
neutrinos are  monochromatic with $E_\nu$=0.861 MeV.

According to  Standard Solar Model (SSM) calculations, pp neutrinos are 
produced at  distance $r$ from the center such that
$r/r_\odot < 0.3$, whereas $^7$Be production is 
peaked much closer to the solar center, at $r/r_\odot \simeq 1/20$, so that 
different neutrinos probe different portions of the solar interior.

The next generation of solar neutrino experiments (e.g.  Hellaz, Borexino 
and Lens) points toward a neutrino spectroscopy, which will elucidate the 
detailed mechanism of energy production in the sun.   

\subsection{Neutrinos as solar thermometers}
\label{subsectherm}

The central temperature $T$ of the  sun is a nice example of a physical 
quantity which can be determined by means of solar neutrino detection, 
provided  that the relevant nuclear physics  is known (and 
neutrino properties are also known).

SSM   calculations predict $T$ with an accuracy of 1\% or even 
better. In order to appreciate such a result, let us remind that the 
central temperature of Earth is known with an accuracy of about 20\%.  
However, let us remind that this is a theoretical prediction which, as any 
result in physics, demands observational evidence.

The fluxes  of $^8$B and $^7$Be neutrinos are given by:
\begin{eqnarray}
\label{eq2a}
	\Phi(B) &=& c_B  S_{17} \frac{S_{34}}{\sqrt{S_{33}}} T^{20} \\
\label{eq2b}
 \Phi(Be) &=& c_{Be} \frac{S_{34}}{\sqrt{S_{33}}} T^{10} 
\end{eqnarray}
where  $S_{ij}$ are the low energy astrophysical factors for nuclear reactions 
between nuclei with atomic mass numbers i and j, $c_B$ and $c_{Be}$ are well 
determined constants.

The high powers of $T$ in the above equations imply that the meaured 
neutrino fluxes are strongly sensitive to $T$, i.e. $^7$Be and $^8$B neutrinos in 
principle are good thermometers for the  innermost part of the sun. On the 
other hand, the relevant  nuclear physics has to be known, which justifies 
the present theoretical and experimental efforts for better determinations 
of the $S_{ij}$.

The result of Superkamiokande \cite{gf:sk} 
can be used for determining a lower limit to $T$. In fact, the
observed flux $\Phi(B)_{obs}=2.44\pm0.10\cdot10^6$ cm$^{-2}$ s$^{-1}$
is a lower limit to the produced flux $\Phi(B)$, since some of the
produced $\nu_e$ can transform into other neutrinos ($\nu_\mu, \nu_\tau$)
with a much smaller detection cross section (by definition
this latter vanishes for oscillation into sterile neutrinos).
By using $\Phi(B)_{obs}\leq \Phi(B)$ together with eq. (\ref{eq2a})
and for the largest (smallest) values of $S_{17} $ and $S_{34}$ 
($S_{33})$ one gets: $T \geq 1.49\cdot 10^7$ K, a value
within five percent from the most recent
SSM estimate, $T_{SSM}=1.57 \cdot 10^7$ K) \cite{gf:BBP98}.

We consider this as one of the successes of the SSM. Let us
observe however that this approach cannot lead to
a measurement of $T$, unless the fate of $\nu_e$ is known,
i.e. the oscillation parameters are determined.

One can conceive a measurement of the central temperature
 which is independent on the oscillation
mechanism. In this direction, Bahcall \cite{gf:broad} has
proposed to measure the difference in average energy between the neutrino
line produced by $^7$Be electron capture in the solar
interior and the corresponding neutrino line produced in terrestrial
laboratory. The high temperatures in the center of the sun cause an everage
energy shift of 1.3 KeV and broaden the line asymmetrically (FWHM=1.6 KeV).

Experimentally this is an extremely  difficult task, in particular
if one aims at  a few percent accuracy  on $T$.
It shows however one of the aspects why detection
of $^7$Be neutrinos is particulary interesting.
In fact  there is a big effort to measure the $^7$Be neutrino
flux  with Borexino, a detector which is being built
by an international collaboration 
at Laboratori Nazionali del Gran Sasso \cite{gf:borexino}.
 
\subsection{Neutrinos from Stellar collapse}
\label{subseccollapse}

As well known, in a stellar collapse leading to Supernova-II explosion most 
of the energy, $E_{SN}\approx 5\cdot 10^{46}$ J, is carried out 
by neutrinos, with average energy
of about 12 MeV, in a  few seconds.
In fact, the detection of a few (anti)neutrinos from SN-1987A 
in the Large Magellan Cloud by Kamiokande and IMB detectors
\cite{gf:SN1987A} opened a completely new field of neutrino astronomy.

The number of detected neutrinos, as well as their energy, was
enough to  show, at a semiquantitative level, that stellar collapse
is essentially understood, see e.g. \cite{gf:loredo}. 
On the other hand, the mechanism leading to the explosion after collapse is 
still a mistery, see e.g.  \cite{gf:koshiba}. 

In this respect it is worth observing that detectors have advanced
substantially. As an example,  for a Supernova-II
at the center of the Galaxy  Superkamiokande will collect
some  5000 events, of which 
1000 already
in the first second after collapse \cite{gf:barrowsa,gf:acerbi}.
This clearly implies that neutrino emission can be followed
quite accurately. In particular, processes with a few millisecond
time scale, i.e. the typical scale of neutron stars ($t=1/\sqrt{G\rho}$)
can be studied. In addition, a detailed neutrino energy spectrum 
can be determined. All this should allow for a basic understanding
of the process, and also clarify between different types of remnants
(e.g. neutron stars, quark stars, black holes and other exotic objects).  
This really will  open, from the observational point of view, 
a completely new field of physics, where nuclear physics is deeply
involved, see e.g. the contributions to this workshop by Drago and
Alberico. 

\subsection{A neutrino map of the Galaxy}
\label{subsecmap}

It is interesting to compare the energy $E$ and neutrinos $N$
released from a stellar 
collapse and from all stars in our galaxy.
For a Supernova one has:
\begin{equation}
\label{eqsn}
E_{SN}\approx 5\cdot 10^{46} {\mbox{J}}\quad ,
\quad N_{SN}\approx 3\cdot 10^{58}
\end{equation}

The luminosity of the Galaxy is about $10^{11} L_\odot=4\cdot 10^{37}$ J/s.
Most of the energy is carried out by photons (not by neutrinos)
and it is produced   through $4p+2e^+\rightarrow ^4He+2\nu +27$ MeV,
i.e. a neutrino is produced for each 13 MeV of electromagnetic energy,
so that the neutrino production rate is about $2\cdot 10^{49}$/s

Roughly, one estimates that there is a Supernova every 30 years in the Galaxy,
see \cite{gf:araa}. In the same time all stars in the  Galaxy have produced:
\begin{equation}
\label{eqstars}
E_{stars} \approx 4\cdot 10^{46} {\mbox{J}}
\quad , \quad N_{stars}= 2\cdot 10^{58}
\end{equation}

There are two curious aspects when comparing 
eq.  (\ref{eqsn}) and eq. (\ref{eqstars}):\\
\noindent
i)on the average,  the contribution of supernovae
to galactic energy production
equals that from all other stars.\\ 
\noindent
ii) again on the average, the contribution of supernovae to
neutrino production in the universe equals that from all other stars.

\begin{figure}[ht]
\centerline{
\epsfysize=10cm \epsfxsize=10cm \epsfbox{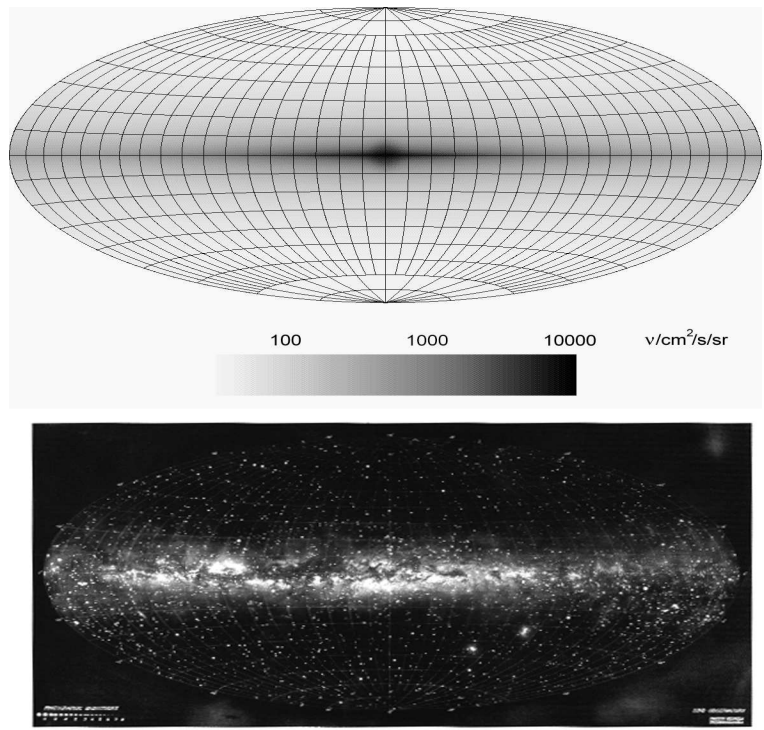}  
}
\caption[ccc]{The Galaxy  seen in neutrinos (top) and in the visible light
(bottom)}
\label{figgalaxy}
\end{figure}

The detection of stellar neutrinos in our galaxy is beyond
present experimental possibility. It is however very interesting
since, due to the long neutrino mean free path, it could provide
an unobscured map of the galaxy, see fig. \ref{figgalaxy} \cite{gf:galaxy}.

\section{Neutrinos as probes of physics beyond the standard
model of electroweak interactions} 

In the Minimal Standard Model of Electroweak interactions ($MSM_{EW}$) 
neutrinos are massless, thus stable and with vanishing electro-magnetic 
moments.

Actually, the deficit of neutrinos reported by all five solar neutrino 
experiments performed so far  strongly points towards some neutrino 
property beyond the $MSM_{EW}$. More importantly, the experiments look in 
contradiction among themselves  unless some non-standard neutrino property 
is advocated, see e.g. refs. \cite{gf:how,gf:where}.
Furthermore, the recent results of Superkamiokande on 
atmospheric neutrinos, confirming  with much higher statistics and better 
particle discrimination  the indications of previous experiments, provide 
additional evidence towards some non standard neutrino property.

The simplest (although not unique) solution of the solar and atmospheric 
neutrino puzzles is in terms of  oscillations  due to a mass difference  
$\Delta m$ among the mass eigenstates, which are superposition of 
weak-flavour eigenstates.
If this mechanism is correct, the real problem now is to determine the 
neutrino mass matrix. This of course is essential for extending  the 
$MSM_{EW}$ and it is also important for understanding the mass generation of 
matter fields. More generally this study is also significant to understand 
which contribution, if any,  to the dark matter in the universe is  due to 
neutrinos.

In the near future neutrino oscillations will be studied with a variety of 
approaches. Long baseline (e.g. CERN to Gran Sasso) accelerator 
experiments are planned and a new generation of refined solar neutrino 
detectors is advancing. All these experiments, however, are sensitive to 
mass difference and cannot determine the absolute mass scale.

In this respect measurements of neutrino masses, e.g. by Tritium beta 
decay and $\beta\beta 0 \nu$  decay (this latter for Majorana neutrino 
only) are extremely important. We recall that if neutrino oscillations 
really occur, these experiments are sensitive to all the neutrino masses, 
and not only to the lightest one \cite{gf:barger}.

\section{The role of nuclear physics}
\label{secrole}

Nuclear physics has an important role in the program outlined above, and 
the theoretical as well as the experimental approach are both important. 
Roughly, one has three kind of problems:

\subsection{Determination of the production cross sections}
\label{subsecprod}

As an example, the  reaction which is at the starting point of the  pp 
chain:
\begin{equation}
\label{eq4}
p+p \rightarrow d+ e^+ + \nu_e
\end{equation}
has a cross section so small that it cannot be measured in the laboratory 
and it has to be determined theoretically.

The  calculated cross section  should now be accurate to the percent  
level \cite{gf:A98,gf:nacre},
 the most recent  and possibily precise determination
 \cite{gf:rosati} being  
produced  by a large international collaboration which  includes
 members of the  MURST project. 

The value of the cross section is
 clearly relevant for determining the 
solar temperature in the energy production zone. The observed solar 
luminosity essentially fixes the pp reaction rate in the sun, which 
depends both on the cross section and on the energy distribution of the 
colliding protons.  From numerical simulations, see e.g. \cite{gf:report},
one has  $T \propto S_{11} ^{-0.1}$, so that a 1\% error on $S_{11}$
translates into a 0.1\% 
error on the temperature, a small however non negligible contribution to  
the present total uncertainty of $T$.

We remark that there are reactions which are
 important for neutrino production,  
although neutrinos are not directly involved.
As an example, let us mention the $^3He+ ^3He \rightarrow ^4He +2p$
and $^3He+ ^4He\rightarrow ^7Be + \gamma$ reactions. As clear from  
Fig. \ref{figpp} and from 
Eqs. (\ref{eq2a},\ref{eq2b}), 
the competition between the two rates determines the flux of $^7$Be 
and $^8$B neutrinos. This is why these reactions are  being  extensively 
studied, both theoretically and experimentally,
 see \cite{gf:A98,gf:nacre}.

\subsection{Determination of the absorption cross section}
\label{subsecdet}

So far, all but one solar neutrino experiments use nuclear targets for 
neutrino detection. With respect to $\nu-e$ scattering,  one loses the 
directionality of the process, however the much larger neutrino-nucleus 
cross sections allow for  larger statistics, a point which is always 
relevant  in solar neutrino physics where one typically collects a few 
hundred events in several years of data  taking. 
Among future projects, let us remind the following reactions, as a couple 
of examples: 
\begin{eqnarray}
 \nu_e + ^{40}Ar  & \rightarrow & ^{40}K + e^- 
\quad {\mbox{(Icarus)\cite{gf:icarus}}} \nonumber \\ 
 \nu_e + ^{160}Gd & \rightarrow & ^{160}Tb +e^- 
\quad {\mbox{(Lens)\cite{gf:lens}}} \nonumber
\end{eqnarray}

The relevant nuclear cross sections are needed, firstly in order to 
fix the size of the apparatus and later in order to intepret the data.

As shown by Gallex and Sage, the apparatus can be calibrated 
by using a suitable  neutrino source and thus the detection 
cross section can be 
determined {\em a posteriori} \cite{gf:calibration}.
 However, it has to be known reliably  in advance  
in order that the the size of the apparatus can be correctly determined, 
i.e. one cannot built a  multi million dollar solar neutrino experiment 
and finally discover that the cross section is too small for solar 
neutrino detection!

\subsection{Determination of the nuclear matrix elements relevant for the 
measurement of neutrino mass}

Neutrinoless double beta decay ($\beta \beta 0 \nu$)
is presently the best tool for determining 
Majorana neutrino masses. Actually the upper bound derived by this 
approach, $ m\lapprox 0.5$ eV, is an order of magnitude smaller that that from 
Tritium beta decay experiments \cite{gf:rpp}.
The interpretation of $\beta\beta 0\nu$ results, however, relies on the 
  calculated nuclear matrix elements, see \cite{gf:klapdor}.

\section{Activity within the network}
\label{secactivity}

The network is active on all the items mentioned in the  previous section. 
Recent activity is exemplified in the four boxes, each containing 
information on selected papers pertinent to this field of physics.

The Milan node, coordinated by  R Broglia, has been  
working in close connection with the experimental groups in the same 
university, which has a very important tradition in experimental neutrino 
physics.   In fact, Broglia et al \cite{gf:brogliana} provided the first estimate of the 
neutrino cross section for $^{23}$Na, which is a crucial ingredient for a 
planned solar neutrino experiment by Fiorini et al. \cite{gf:fiorini}.
 They also  produced  a 
more accurate  estimate for the neutrino cross section on $^{40}$Ar, a result 
important for Icarus \cite{gf:brogliaar}. The new calculation of electron capture of 
$^{123}$Te \cite{gf:brogliate}
helped resolve a discrepancy between the experimental 
results of  Fiorini et al.  \cite{gf:fiorini2}
and the previous ones by another group.

The Pisa group, coordinated by S. Rosati, has  developped over the years 
more and more refined calculation methods 
for few body systems, see \cite{gf:rosatia,gf:rosatib}, which are 
becoming important for applications to nuclear astrophysics, a field where 
the activity of the group will concentrate in the near  future.  In this 
respect, it is worth mentioning the contribution \cite{gf:rosati} 
 of this group to the 
study of the weak capture of protons by protons, a process which 
importance has been previously mentioned.

The same process has been studied by the Ferrara group, coordinated by G. 
Fiorentini, within a completely different approach. We  already remarked 
that it is impossible to measure the
cross section in the laboratory. However one can get observational 
constraints on its value by exploiting helioseismic data \cite{gf:spp}. 

Nuclear reaction rates in stars are also investigated by the group of 
``Torino-Politecnico", coordinated by P. Quarati. In particular, they have 
considered the possible modification of reaction rates, due to deviations 
from the Boltzmann  statistics, in the framework of the extended Tsallis 
statistics \cite{gf:quaratia,gf:quaratib}.
 This hypothesis cannot be discarded a priori, in the presence 
of long range forces. In a paper coauthored with the Ferrara group 
\cite{gf:eliosstat}, again 
by exploiting helioseismic constraints, it has been shown that such 
deviations, if they exist, are extremely tiny.

Clearly there is a strong overlap of scientific interests among the 
different groups and similar problems, e.g. the pp reaction, are 
approached with complementary methods. There is more joint work than the 
number of coauthored papers would suggest. We are confident that several 
presently informal collaborations will bring in joint papers in the near 
future, see the next section.

\vspace{-2cm}

\framebox(346,520)[h]{
\parbox[t]{4.3in}{
{\scriptsize
{\large {\bf{ The solar neutrino capture corss section for $^{23}$Na}}  }\\
{\large {\bf{  }}} W.E.Ormand, P.M. Pizzochero, P.F. Bortignon, and R.A. Broglia.\\
{\large {\bf{  }}} Phys. Lett. B308 (1993) 207\\
    A coincidence (eletron, gamma-ray) experiment designed to identify one or more components 
    of the solar neutrino spectrum has been proposed based on a large array of NaBr detectors. 
    In support of the design of this detector , we calculate the solar neutrino absorption rate for 
    n+$^{23}$Na$\rightarrow ^{ 23}$Mg+e within the Standard Solar Model making use of both experimental and 
    theoretical data for the structure of the two nuclei involed. It is found that the inclusion of 
    excited states in $^{23}$Mg enhances the absorption cross section by $\simeq 30\%$, with approximately 
    on third of this enahcement coming from excited states for which experimental ata does not exist. 
    The solar neutrino absorption rate is calculated to e 3.5$\pm$1.3 SNU, which ammounts to aboutn 
    one count every six days for the proposed detector.\\
~\\
{\large {\bf{ Neutrino Capture Cross Sections for $^{40}$Ar and beta-decay of $^{40}$Ti }}}\\
{\large {\bf{  }}} W.E.Ormand, P.M. Pizzochero, P.F. Bortignon, and R.A. Broglia. \\
{\large {\bf{  }}} Phys. Lett. B345 (1995) 343\\
    Shell-model calculations of solar neutrino absorption cross sections for $^{40}$Ar, the
    proposed component of the ICARUS detector, are presented. It is found that low-lying
    Gamow-Teller transitions lead to a significant enhancement of the absorption rate over
    that expected from the Fermi transition between the isobaric analog states, leading to an
    overall absorption rate of 6.7 SNU. We also note that the pertinent Gamow-Teller
    transitions in $^{40}$Ar are experimentally accessible from the $\beta$-decay of the  mirror 
    nucleus 40Ti. Predictions for the branching ratios to states in  $^{40}$Sc are presented, 
    and the theoretical halflife of 53 ms is found to be in good  agreement with the experimental 
    value of 56$^{+18} _{-12}$ ms.\\
~\\
{\large {\bf{ Competition Between Particle-Hole and Particle-Particle Correlations 
in Forbidden Electron Capture: the Case of $^{123}$Te }}} \\
{\large {\bf{  }}} M. Bianchetti, M. R. Quaglia,G. Col\`o, P.M. Pizzochero, R.A. Broglia  and P.F. Bortignon \\
{\large {\bf{  }}} Phys. Rev. C 56 (1997) R1676 \\
    The K-electron capture half-life of $^{123}$Te has been recently measured to be $K_{exp}=2.4 \cdot 10^{19}$yr, 
    and constitutes the longest half-life ever measured in a single b transition of any nuclear species. 
    We have calculated this second unique forbidden transition within the framework of the 
    proton-neutron  quasi-particle random  phase approximation, making use of Skyrme-type effective
    interactions. A strong cancellation effect between particle-hole and particle-particle correlations is found.
   The model, without any renormalization of the force, provides a lower limit for the K-electron
    capture half-life of $\simeq 10^{17}$yr, which unambiguously rules out the old experimental values 
    of $10^{13}-10^{14}$ yr. A few percent increase of the particle-particle matrix elements of the Skyrme 
    interaction allows to reproduce the experimental findings.\\
}}
}
\vspace{1 cm}
\centerline{Box 1: Milan results}

\framebox(346,520)[h]{
\parbox[t]{4.3in}{
{\footnotesize
{\large {\bf{ Weak capture of protons by protons }}}\\
{\large {\bf{ }}} R. Schiavilla, V. G. J. Stoks, W. Gloeckle, 
H. Kamada, A. Nogga, J. Carlson, R.Machleidt, V. R. Pandharipande, 
R. B. Wiringa, A. Kievsky, S. Rosati, M. Viviani\\
{\large {\bf{ }}} Phys.Rev. C58 (1998) 1263\\
    The cross section for the proton weak capture reaction $^1$H(p,e$^+$+$\nu_e$)$^2$H is
    calculated with wave functions obtained from a number of modern, realistic
    high-precision interactions. To minimize the uncertainty in the axial two-body current
    operator, its matrix element has been adjusted to reproduce the measured Gamow-Teller
    matrix element of tritium $\beta$ decay in model calculations using trinucleon wave
    functions from these interactions. A thorough analysis of the ambiguities that this
    procedure introduces in evaluating the two-body current contribution to the pp capture is
    given. Its inherent model dependence is in fact found to be very weak. The overlap
    integral L2(E=0) for the pp capture is predicted to be in the range 7.05--7.06,
    including the axial two-body current contribution, for all interactions considered.\\
~\\
{\large {\bf{ Neutron-$^3$H and Proton-$^3$He Zero Energy Scattering }}} \\
{\large {\bf{ }}} M. Viviani, S. Rosati, A. Kievsky\\
{\large {\bf{ }}} Phys.Rev.Lett. 81 (1998) 1580-1583 \\
    The Kohn variational principle and the (correlated) Hyperspherical Harmonics technique
    are applied to study the n-$^3$H and p-$^3$He scattering at zero energy. Predictions for the
    singlet and triplet scattering lengths are obtained for non-relativistic nuclear Hamiltonians
    including two- and three-body potentials. The calculated n-$^3$H total cross section agrees
    well with the measured value, while some small discrepancy is found for the coherent
    scattering length. For the p-$^3$He channel, the calculated scattering lengths are in
    reasonable agreement with the values extrapolated from the measurements made above 1
    MeV.\\
~\\
{\large {\bf{Possible three-nucleon force effects in D-P scattering at low energies }}} \\
{\large {\bf{ }}}C. R. Brune, W. H. Geist, H. J. Karwowski, E. J. Ludwig, K. D. Veal, M. H. Wood,
 A. Kievsky, S. Rosati, M. Viviani \\
{\large {\bf{ }}}Phys.Lett. B428 (1998) 13-17 \\
    We present measurements of the analyzing powers Ay and iT11 for proton-deuteron
    scattering at Ecm=432 keV. Calculations using a realistic nucleon-nucleon potential
    (Argonne V18) are found to underpredict both analyzing powers by 40. The inclusion of
    the Urbana three-nucleon interaction does not significantly modify the calculated
    analyzing powers. Due to its short range, it is difficult for this three-nucleon interaction to
    affect Ay and iT11 at this low energy. The origin of the discrepancy remains an open
    question.\\
}}
}
\vspace{1 cm}
\centerline{Box 2: Pisa results}


\framebox(346,520)[h]{
\parbox[t]{4.3in}{
{\footnotesize
{\large {\bf{ Superkamiokande and solar antineutrinos }}}\\
{\large {\bf{ }}} G. Fiorentini ,M. Moretti , F. L. Villante \\
{\large {\bf{ }}} Phys.Lett. B413 (1997) 378-381. \\
 We propose to exploit the angular distribution of the positrons
emitted in the inverse beta decay to extract a possible antineutrino
signal from the Superkamiokande background. 
From the statistics collected in just 101.9 days one obtains a model independent
 upper bound
on the antineutrino flux (for energy greater than 8.3 MeV) 
$\Phi < 9\cdot10^4$ cm$^{-2}$ s$^{-1}$
at the 95\% C.L. By assuming the same energy spectrum 
as for the $^8$B neutrinos, the
95\% C.L. bound is $\Phi < 6\cdot10^4 cm^-2 s^-1$. 
Within three years of data taking the sensitivity
to neutrino-antineutrino transition probability 
will reach the 1\% level, thus providing
a stringent test of hybrid oscillation models.\\
~\\
{\large {\bf{Helioseismology and  p+p $\rightarrow$ d + e$^+$ + $\nu _e$ in the
sun  }}} \\
{\large {\bf{ }}} S. Degl'Innocenti, G. Fiorentini, B. Ricci \\
{\large {\bf{ }}} Phys.Lett. B416 (1998) 365-368 \\
By using a phenomenological field theory of nucleon-nucleon interactions, 
Oberhummer 
et al. found a cross section of p+p $\rightarrow$ d + e$^+$ + $\nu _e$ 
about 2.9 times that given 
by the potential approach and adopted in Standard Solar Model calculations. 
We show 
that a solar model with $S=2.9 S_{SSM}$ is inconsistent with helioseismic 
data, the
difference between model predictions and helioseismic 
determinations being typically a 
factor ten larger than estimated uncertainties. We also 
show that, according to 
helioseismology, $S$ cannot differ from $S_{SSM}$ by more than 15\%.\\
~\\
{\large {\bf{Bounds on $hep$ neutrinos}}} \\
{\large {\bf{ }}} G. Fiorentini, V. Berezinsky, S. Degl'Innocenti, B. Ricci \\
{\large {\bf{ }}} astro-ph/9810083, to appear on  Phys. Lett. B (1998) \\
The excess of highest energy solar-neutrino 
events recently observed by Superkamiokande 
can be in principle explained by anomalously 
high $hep$-neutrino flux $\Phi_{\nu}(hep)$. 
Without using SSM calculations, from the solar 
luminosity constraint we derive that  
$\Phi_\nu(hep)/S_{13}$ cannot exceed the SSM 
estimate by more than a factor three. If one makes  
the additional hypothesis that $hep$ neutrino  
production occurs where the $^3$He 
concentration is at equilibrium, helioseismology 
gives an upper bound which is (less then)
two times the SSM prediction. We 
argue that the anomalous $hep$-neutrino flux of order of 
that observed by Superkamiokande cannot 
be explained by astrophysics, but rather by a large 
production cross-section.\\
}}
}
\vspace{1 cm}
\centerline{Box 2: Ferrara results}



\framebox(346,520)[h]{
\parbox[t]{4.3in}{
{\scriptsize
{\large {\bf{  Anomalous diffusion modifies solar neutrino fluxes }}} \\
{\large {\bf{ }}} G. Kaniadakis, A. Lavagno, M. Lissia, P. Quarati\\
{\large {\bf{ }}} astro-ph/9710173, to appear on Physica A (1998)\\
     Density and temperature conditions in the solar core suggest that the microscopic
    diffusion of electrons and ions could be nonstandard: Diffusion and friction coefficients 
    are energy dependent, collisions are not two-body processes and retain memory beyond 
     the single scattering event. A direct consequence of nonstandard diffusion is that the 
     equilibrium energy distribution of particles departs from the Maxwellian one 
    (tails goes to zero more slowly or faster than exponentially) modifying the reaction rates. 
    This effect is qualitatively different from temperature and/or composition modification:
    Small changes in the number of particles in the distribution tails can strongly modify the
    rates without affecting bulk properties, such as the sound speed or hydrostatic
    equilibrium, which depend on the mean values from the distribution. 
    This mechanism can considerably increase the range of predictions for the neutrino
    fluxes allowed by the current experimental values (cross sections and solar properties)
    and can be used to reduce the discrepancy between these predictions and the solar
    neutrino experiments.\\
~\\
{\large {\bf{ Helioseismology can test the Maxwell-Boltzmann distribution }}} \\
{\large {\bf{ }}} S. Degl'Innocenti, G. Fiorentini, M. Lissia, P. Quarati, B. Ricci \\
 {\large {\bf{ }}} astro-ph/9807078, to appear on  Phys. Lett. B \\
    Nuclear reactions in stars occur between nuclei in the high-energy tail of the energy
    distribution and are sensitive to possible deviations from the standard equilibrium
    thermal-energy distribution. We are able to derive strong constraints on such deviations
    by using the detailed helioseismic information of the solar structure. If a small deviation is
    parameterized with a factor exp{-$\delta*(E/kT)^2$}, we find that delta should lie between
    -0.005 and +0.002. However, even values of delta as small as 0.003 would still give
    important effects on the neutrino fluxes.\\
~\\
{\large {\bf{  Non-Markovian effects in the solar neutrino problem }}} \\
{\large {\bf{ }}} G. Gervino, G. Kaniadakis, A. Lavagno, M. Lissia, P. Quarati \\
{\large {\bf{ }}} physics/9809001,to appear in the Proceedings of Nuclei in the Cosmos V (1998)\\
    The solar core, because of its density and temperature, is not a weakly-interacting or a
    high-temperature plasma. Collective effects have time scales comparable to the average
    time between collisions, and the microfield distribution influences the particle dynamics.
    In this conditions ion and electron diffusion is a non-Markovian process, memory effects
    are present and the equilibrium statistical distribution function differs from the
    Maxwellian one. We show that, even if the deviations from the standard velocity
    distribution that are compatible with our present knowledge of the solar interior are small,
    they are sufficient to sensibly modify the sub-barrier nuclear reaction rates. The
    consequent changes of the neutrino fluxes are comparable to the flux deficits that
    constitute the solar neutrino problem.\\
}}
}
\vspace{1 cm}
\centerline{Box 4: Turin results}

\section{Hot topics}

In this section we consider a few points, which look presently as 
particularly interesting and which will be investigated by our 
collaboration, in the near future.

\subsection{ $hep$ neutrinos}

They are produced by means of the reaction,
\begin{equation}
\label{eqhep}
 		p+ ^3He  \rightarrow ^4He  + e^+ + \nu_e \quad ,
\end{equation}
a very marginal branch of the pp-chain, $\Phi(hep)/\Phi(tot) \approx10^{-8}$,
 which 
gives the highest energy solar neutrinos (E$_{max}$ =18.8 MeV).
In the last few months, these rare neutrinos have become particularly 
interesting in view of the surprising result of Superkamiokande
\cite{gf:sk}, which 
reported an excess of events near and beyond the end point of the $^8$B 
spectrum.  This result might
 be explained \cite{gf:BK98} by an anomalously high 
$hep$ flux, 
\begin{equation}
\label{eqhep2}
\Phi(hep) \simeq 30 \Phi(hep)_{SSM}
\end{equation}
where $\Phi(hep)_{SSM}=2.1\cdot 10^3$cm$ ^{-2}$ s$^{-1}$ is 
 the SSM prediction \cite{gf:BBP98}. 

This  result has to be taken  with some reservation. The experimental 
indication is not that strong, and actually needs confirmation. Within one 
year Superkamiokande will have collected enough data to establish beyond 
any statistical doubt if the excess of high energy events is there. 
Furthermore, more or less in the same time data  from SNO should be 
available. These are particularly interesting, since the neutrino 
detection is based on a completely different method.

Let us observe that for the first time we have an 
excess, and not a deficit, of solar neutrinos!  Beyond this possibly 
amusing point, one has to remark the substantial progress of our 
experimental colleagues. Solar neutrino physics has really advanced to a 
stage such that even the rarest  branch of the pp chain  has possibily 
been detected.

The measurement of $\Phi(hep)$ is really a benchmark for testing our present 
knowledge of stellar interiors, if the zero energy astrophysical S-factor 
of reaction (\ref{eqhep}) has been correctly calculated. 

In fact one can derive on very general grounds upper bounds on the ratio  
$\Phi(hep)/S_{13}$ \cite{gf:hep}:\\

\noindent
i) From the luminosity constraint, i.e. the fact that the presently 
observed solar luminosity equals the nuclear power  presently generated in 
the solar interior, one has:
\begin{equation}  
\Phi(hep)/S_{13} \leq   3  \Phi(hep)_{SSM}/S_{13 \, SSM} \quad .
\end{equation}

\noindent
ii) If one assumes that $hep$ neutrinos are produced in a region where $^3$He 
abundance has reached local equilibrium -- an assumption which is actually 
a result in many standard and non standard solar model calculations -- one 
gets a stronger bound by using helioseismology:
\begin{equation}
\Phi(hep)/S_{13} \leq   1.7  \Phi(hep)_{SSM}/S_{13 \, SSM} \quad .
\end{equation}

In other words, if $\Phi(hep) \simeq  30 \Phi(hep)_{SSM}$ is confirmed and the 
presentvalue $S_{13\,SSM}=(2.3 \pm 0.9) \cdot 10^{-20}$ KeVb 
\cite{gf:SC92,gf:A98}
is confirmed, then one should abandon the 
common view of the solar interior, which should be in a state dramatically 
out of equilibrium conditions.

All this shows the relevance of an accurate calculation of $S_{13}$, a rather 
difficult task indeed. With respect to the intitial estimate by Salpeter 
\cite{gf:S52}, the presently accepted value is smaller by two orders of magnitude. 
This results from selection rules as well as from subtle cancellations, as 
emphasized in \cite{gf:SC92}. The authors of \cite{gf:SC92}
 are very careful in 
estimating an uncertainty of at least a factor two in the recommended 
value.  In view of the present situation, a more refined examination, 
including fully state-of-the art methods of few body physics is clearly 
needed, and the Pisa group has all the technology which is needed for this 
difficult goal.

\subsection{Be-e-p neutrinos}

In addition to neutrinos from $^8$B beta decay, a few neutrinos can also 
result from electron capture, $e^- +^8B \rightarrow \alpha + \alpha + \nu_e$.
 Their energy is 
clearly  $2m_e$ above the end point of the $^8$B
 decay, so that they are in the 
region of the event excess reported by Superkamiokande. 
Evaluation of the flux of these neutrinos is thus clearly interesting.
One has to  remind that in the calculation  one has to take into account 
the full reaction which yields these neutrinos:
\begin{equation}
\label{eqbep}
                 ^7Be+e^-+p \rightarrow \alpha + \alpha + \nu_e
\end{equation}
which justifies the nickname ``Be-e-p". 
The calculation is in progress, within a collaboration between Milan and 
Ferrara.

\subsection{p-e-p neutrinos}

These too are (relatively) rare neutrinos, from:
\begin{equation}
\label{eqpep}
     p+e+p \rightarrow d + \nu_e
\end{equation}
the  estimated flux being about one percent of the total neutrino flux.

The ratio $R = \Phi(pep)/\Phi(pp)$ is an important  quantity,
 for discriminating 
among several proposed solutions of the solar neutrino puzzle.  
The point 
is that $R$ should be predicted very accurately, since
 several uncertainties, originating from both 
nuclear physics and astrophysics, cancel in the ratio.
On the other hand, there are oscillation 
mechanisms which predict strongly different oscillation probabilities for 
pp and pep neutrinos \cite{gf:CAL95}. Furthermore, the ratio  $R$ will 
be measured in future generation solar neutrino experiments, e.g. Hellaz.

So far,  $R$ has been calculated  many years ago only in \cite{gf:BM69}, 
using several approximations. On the other hand, such a few body problem 
is within the possibility of  an ``ab initio" calculation. Due to the 
importance of this quantity, the Pisa and Ferrara nodes are planning to 
produce a new estimate in the near future.

\subsection{ LENS}

\begin{figure}
\centerline{
\epsfysize=12cm \epsfbox{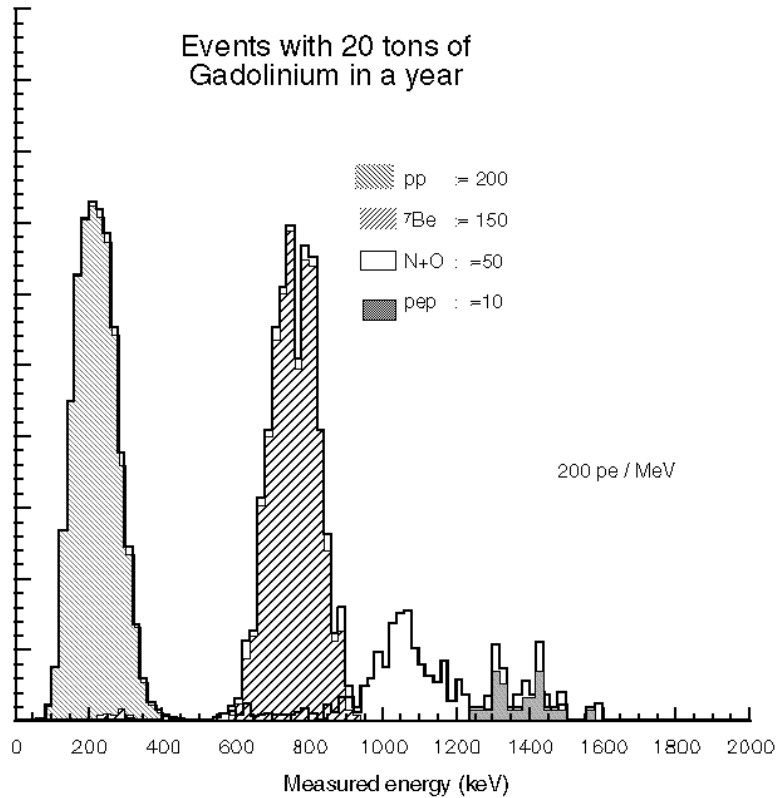}  
}
\caption[ ]{ Spectrum of expected events in LENS experiment. }
\label{figlens}
\end{figure}

A new   solar neutrino experiment has been proposed recently   by 
Raghavan \cite{gf:lens}. 
The experimental aim is really a Low Energy Solar Neutrino  
Spectroscopy - hence the acronym LENS - of unprecedented quality, see 
Fig \ref{figlens}.
 The experiment should clearly discriminate between pp and Be 
neutrinos, a feature which is very important again in connection with the 
predictions of different oscillation mechanisms. Concerning Be neutrinos, 
one has to remark the complementarity between this experiment and 
Borexino. Since LENS is sensitive only to $\nu_e$,  whereas the signal in 
Borexino gets contribution from any active neutrino species, the 
comparison between the results of the two experiments should allow to 
extract the signature of active neutrinos other than $\nu_e$, i.e. one can 
exploit the conjunction between the two experiments so as to realize an 
``appearance" experiment. We remak also that LENS  should be able to detect 
neutrinos from the CNO cycle.

The reaction to  be used in Lens is
\begin{equation}
\label{eqlens}
\nu _e + ^{160}Gd \rightarrow ^{160}Tb^\ast + e
\end{equation}
The electron is detected in (delayed) coincidence with the  gammas
resulting, a few nanoseconds after neutrino capture, 
from  $^{160}Tb^\ast$ decay, 
so as to reduce background to acceptable levels.

Of course, the cross section of (\ref{eqlens})  will have to  be measured
 ``a posteriori" with the same detector used for solar neutrinos, by 
irradiating the target with an artificial neutrino source, e.g. the $^{51}$Cr 
source  used by Gallex or a similar and possibily more powerful one. 
However, a knowledge, as good
as possible,  of the cross section  is important  ``a priori" for planning 
the detector and determine its size.  Some indirect experimental data is 
available, by means of a recent study of the  Gd($^3$He,$^3$H)Tb  reaction at 
Osaka. A significant theoretical effort however is needed in order to 
understand the complicated excitation pattern and the strength of the 
transitions to different excited states, an activity which looks well on 
the research lines of the Milan group.

\subsection{ LUNA2 at LNGS}

Some  years ago Rolfs and Fiorentini proposed  that INFN exploits the 
underground facilities at Laboratori Nazionali del Gran Sasso to measure 
nuclear cross sections at the low energies of astrophysical interest, in  
an environment naturally shielded against the cosmic radiation, so that 
even very low reaction rates can be discriminated from background, see
box 5.

\framebox(346,520)[h]{
\parbox[t]{4.3in}{
To the President of INFN\\
Prof. Nicola Cabibbo\\
~\\
cc:\\
~\\
To the Director of The Gran Sasso Lab\\
Prof. Enrico Bellotti\\
~\\
 \hfill{ Ferrara 7 January 1991} \\
~\\
Dear President,\\
~\\
We believe that the Gran Sasso Laboratory offers a unique possibility for 
progress in the measurement of low energy nuclear cross sections, which
are relevant for nucleosynthesis in stars and in the early universe, as 
well as for the evaluation of the solar neutrino flux.\\
~\\
In this context, hydrogen burning processes operating in the p-p chain 
(low-mass stars, such as the sun) and in the CNO-cycles (high-mass stars)
are of particular interest. Experimental studies of these processes have
been carried out to energies E far below the respective Coulomb barriers
E$_c$. However, experiments have been performed so far typically at energies
E/E$_c$ greater than 1/20, whereas the stellar burning occurs at 
E/E$_c \approx 1/100$. Thus, the available data have always to be extrapolated
over a relatively wide energy range, leading to substantial uncertainties.

These measurements have been optimized by using the best experimental
techniques available today. However, they are all basically limited
by the effect of cosmic rays. \\
~\\
This problems can be reduced significantly
(about three order of magnitude) by carring out such experiments in an
underground laboratory, such as the LNGS.
~\\
\centerline{ . . . . . . .}
~\\
Sincerely yours with best regards\\
~\\

$\quad$ Claus Rolfs $\quad \quad \quad \quad \quad \quad \quad $ Gianni Fiorentini\\
$\quad$ (Univ. of Bochum) $\quad \quad \quad \quad \quad $ (Univ. of Ferrara)\\
}
}
\vspace{1 cm}
\centerline{Box 5: Excerpt from  the letter sent to the INFN president for 
starting the LUNA experiment.}

INFN  answered quickly, actually on the phone.
A small 30 kV accelerator was installed at LNGS  and in 
this way  the first Laboratory for Underground Nuclear Astrophysics (LUNA) 
was born. LUNA has been  successful \cite{gf:lunaa,gf:lunab}.
As an example, for the 
first time a nuclear reaction, $^3He+^3He \rightarrow ^4He + 2p$,
was measured at the energies relevant for burning in the sun \cite{gf:lunab}. 
Corvisiero et al. \cite{gf:luna2} are now  planning 
LUNA2, a second generation apparatus involving a 200 kV accelerator, which 
should be capable of measuring several other cross sections relevant to H 
burning in the sun via the pp-chain and the CNO cycle,
see table. Furthermore, the experiment can also measure other few body 
reactions at very low energies, so that, e.g.,  electron 
screening of nuclear reactions can be elucidated. This will offer a unique 
opportunity to test the most accurate calculation methods of few body 
nuclear phyisics, so that one can again expect a strong collaboration 
between  the experimental group and our theoretical network.

\begin{table}
\centering
\caption[cc]{The proposed schedule of LUNA2 at LNGS}
\begin{tabular}{|lc|}
\hline
&\\
1999:&Installation and testing of a new 200KV accelerator\\
&\\
2000&$^3He(\alpha,\gamma)^7Be$ \\
&\\
2001&$^7Be(p+\gamma)^8B$ \\
&\\
2002&$^{14}N(p,\gamma)^{16}O$ \\
&\\
\hline
\end{tabular}
\label{tabluna}
\end{table}

\section*{Acknowledgments}

I would like thank particularly Giovanni Corbelli and
Barbara Ricci for their help in preparing this compuscript.
I am grateful to Pierfrancesco Bortignon, Sergio Rosati and
Pierino Quarati, for fruitful discussion
about the activity within the network.
This work was supported by Ministero dell' Universita' e
della Ricerca Scientifica e Tecnologica (MURST).

\end{document}